\documentclass[9pt,sigconf,screen]{acmart} 

\usepackage{xcolor,colortbl}
\usepackage{siunitx}
\usepackage[multiple]{footmisc}
\usepackage{url}
\definecolor{seaborn_muted_green}{HTML}{6ACC64}
\definecolor{seaborn_muted_red}{HTML}{D65F5F}

\AtBeginDocument{%
  \providecommand\BibTeX{{%
    \normalfont B\kern-0.5em{\scshape i\kern-0.25em b}\kern-0.8em\TeX}}}

\copyrightyear{2021} 
\acmYear{2021} 
\setcopyright{acmlicensed}
\acmConference[Middleware '21]{22nd International Middleware Conference}{December 6--10, 2021}{Virtual Event, Canada}
\acmBooktitle{22nd International Middleware Conference (Middleware '21), December 6--10, 2021, Virtual Event, Canada}
\acmPrice{15.00}
\acmDOI{10.1145/3464298.3493399}
\acmISBN{978-1-4503-8534-3/21/12}

\newcommand{\coo}{\ensuremath{\mathrm{CO_2}}}
\newcommand{\cooeq}{\ensuremath{\mathrm{CO_2 eq}}}
\newcommand{\coopkwh}{\si{gCO_2 \per kWh}}

\begin{document}

\title{Let's Wait Awhile: How Temporal Workload Shifting Can Reduce Carbon Emissions in the Cloud}
\renewcommand{\shorttitle}{Let's Wait Awhile: How Temporal Workload Shifting Can Reduce Carbon Emissions in the Cloud}


\author{Philipp Wiesner}
\email{wiesner@tu-berlin.de}
\orcid{1234-5678-9012}
\affiliation{%
  \institution{Technische Universität Berlin}
  \city{Berlin}
  \country{Germany}
}

\author{Ilja Behnke}
\email{i.behnke@tu-berlin.de}
\affiliation{%
  \institution{Technische Universität Berlin}
  \city{Berlin}
  \country{Germany}
}

\author{Dominik Scheinert}
\email{dominik.scheinert@tu-berlin.de}
\affiliation{%
  \institution{Technische Universität Berlin}
  \city{Berlin}
  \country{Germany}
}

\author{Kordian Gontarska}
\email{kordian.gontarska@hpi.de}
\affiliation{%
  \institution{HPI, University of Potsdam}
  \city{Potsdam}
  \country{Germany}
}

\author{Lauritz Thamsen}
\email{lauritz.thamsen@tu-berlin.de}
\affiliation{%
  \institution{Technische Universität Berlin}
  \city{Berlin}
  \country{Germany}
}

\begin{abstract}
Depending on energy sources and demand, the carbon intensity of the public power grid fluctuates over time.
Exploiting this variability is an important factor in reducing the emissions caused by data centers.
However, regional differences in the availability of low-carbon energy sources make it hard to provide general best practices for when to consume electricity.
Moreover, existing research in this domain focuses mostly on carbon-aware workload migration across geo-distributed data centers, or addresses demand response purely from the perspective of power grid stability and costs.

In this paper, we examine the potential impact of shifting computational workloads towards times where the energy supply is expected to be less carbon-intensive.
To this end, we identify characteristics of delay-tolerant workloads and analyze the potential for temporal workload shifting in Germany, Great Britain, France, and California over the year 2020. 
Furthermore, we experimentally evaluate two workload shifting scenarios in a simulation to investigate the influence of time constraints, scheduling strategies, and the accuracy of carbon intensity forecasts.
To accelerate research in the domain of carbon-aware computing and to support the evaluation of novel scheduling algorithms, our simulation framework and datasets are publicly available.

\end{abstract}




\maketitle

\section{Introduction}

\begin{figure}[b!]
    \centering
    \includegraphics[width=0.95\columnwidth]{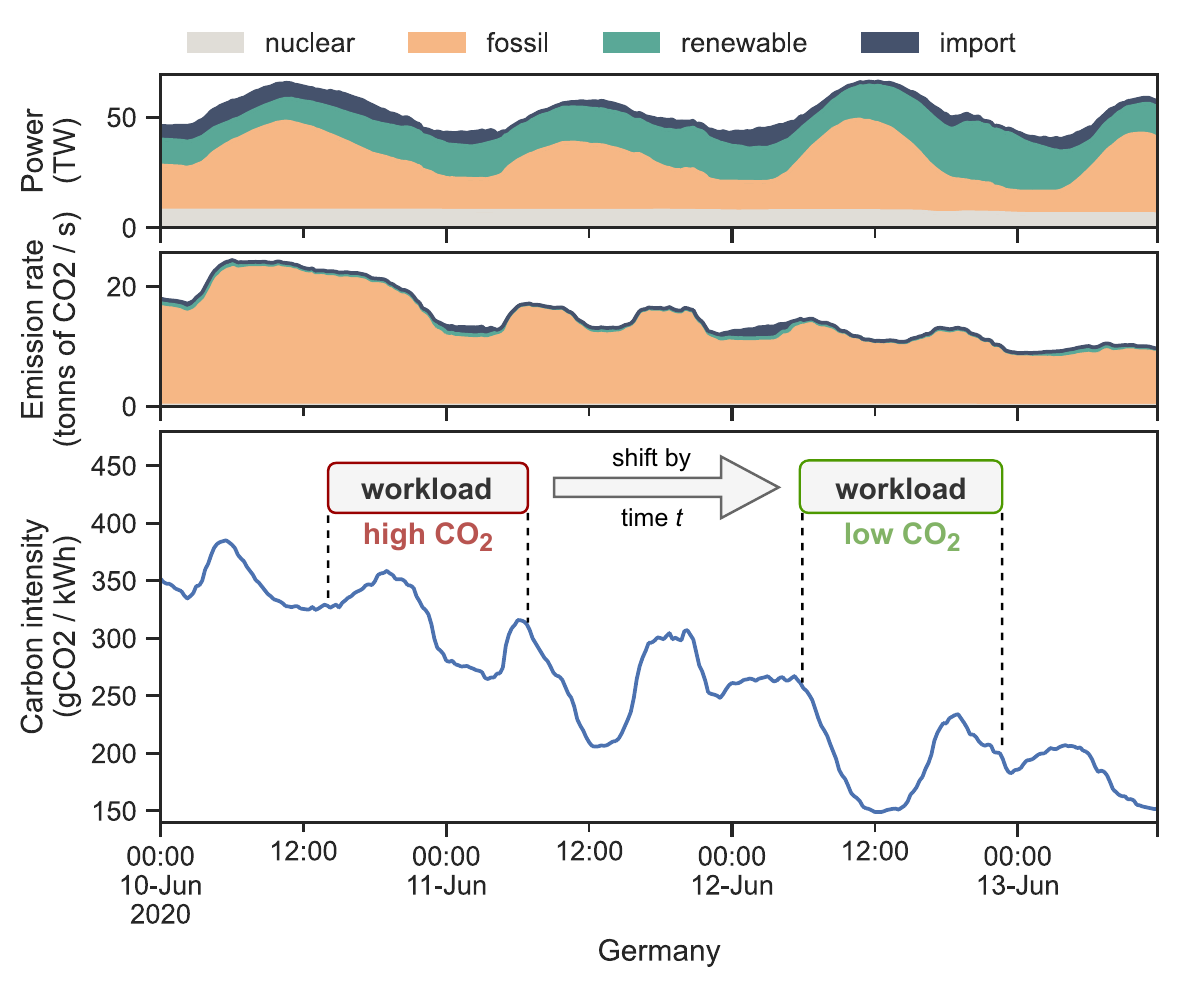}
	\caption{Power consumption, emission rate, and resulting carbon intensity in Germany, June 10-13. Scheduling workloads at times when the carbon intensity is expected to be low, can reduce the carbon footprint of data centers.}
    	\label{fig:fig1}
\end{figure}

Reducing the energy demand of data centers is a major concern of research and industry alike, as it is a key driver of operational expenses and largely determines the carbon footprint of cloud computing.
The extent of these efforts is most evident in the fact that data center energy consumption has grown at a much slower rate over the past decade than previously assumed~\cite{Masanet_RecalibratingGlobalDCEnergyEstimates_2020}.
This success can be attributed to technological advances such as improved processor and storage-drive efficiency on the one side, but even more importantly to the steady shift of cloud computing towards highly energy-optimized hyperscale data centers~\cite{CiscoGlobalCloudIndex_2018} that already account for roughly 50\,\% of all compute instances~\cite{Masanet_RecalibratingGlobalDCEnergyEstimates_2020}.
Despite all the efficiency gains, data centers worldwide consumed an estimated \SI{205}{TWh} of electricity in 2018, which amounts to approximately \SI{1}{\percent} of global energy consumption, and demand is expected to rise further in the future \cite{Masanet_RecalibratingGlobalDCEnergyEstimates_2020}.

IT industry and public cloud providers are pushing towards reducing their impact on the climate, reinforced by a global initiative to implement carbon pricing mechanisms, such as emission trading systems (ETS) or carbon taxes \cite{WorldBank_CarbonPricing_2020}.
However, when targeting the carbon footprint of data centers, not only the amount of energy consumed is important, but also the energy sources.
For example, Google plans to operate their data centers solely on carbon-free energy by 2030 \cite{Google_CarbonFreeBy2030_2020}.
This commitment is much more extensive than what other companies tout as "carbon-free", which often only involves purchasing green power and offsetting their emissions.
True carbon-free operation, on the other hand, is very hard to achieve: Given the variable nature of many renewable energy sources, such as solar and wind, operators must not only invest in energy storage systems, but also manage their demand adaptively to consume energy when and where it is emitting the least \coo
\footnote{Albeit being the most prominent source of global warming, carbon dioxide (\coo) is not the only gas responsible for climate change. Hence, to provide a common scale for describing all greenhouse gases, a popular unit of measurement is the so called \emph{carbon dioxide equivalent}, often abbreviated as \cooeq. For any gas it is defined as the amount of \coo\ that would be needed to warm the earth equivalently. For simplicity, in this article we refer to \cooeq\ when talking about \coo\ or carbon emissions.}.

Energy sources used for electricity production vary highly in different regions, at different seasons, and at different hours of the day. This variability depends on many factors, such as weather and climate, the installed capacity of different energy sources in a region, as well as energy imports from neighboring regions.
The goal of this paper is to investigate the potential impact of shifting delay-tolerant data center workloads towards times where the grid is expected to provide clean energy, as exemplary illustrated in \autoref{fig:fig1}.
To clearly state the \textbf{boundaries} of our research we note that
\begin{itemize}
	\item the aim of this work is not to save energy but to consume energy at times, where it is generated by low-carbon sources.
	\item we aim at exploiting the fluctuation of carbon intensity in the public power grid and do not address the integration of local power generation that provides the data center with its own energy.
	\item we observe the potential of rescheduling on the time dimension. We do not consider any forms of load migration between geo-distributed data centers.
\end{itemize}

Although temporal workload shifting is already finding its way to production environments \cite{Google_SunShinesWindBlows_2020}, existing work in the domain of carbon-aware scheduling mostly focuses on either the integration of renewable on-site or off-site installations \cite{Akoush_FreeLunch_2011, Zhang_GreenWare_2011, Aksanli_GreenEnergyPredictionScheduleBatchServiceJobs_2011, Goiri_MatchingRenewableEnergyGreenDatacenters_2015, Liu_RenewableCoolingAwareWorkloadManagement_2012, Goiri_GreenHadoop_2012, Goiri_ParasolAndGreenSwitch_2013, Liu_RenewableCoolingAwareWorkloadManagement_2012, Li_iSwitch_2012, Dupont_ApplicationControllerForOptimizingRenewableEnergy_2015} or on geo-distributed load migration \cite{Zheng_MitigatingCarbonLoadMigration_2020, Zhou_CarbonAwareLoadBalancingGeoDistributedCloudServices_2013, Moghaddam_CarbonAwareDistributedCloudGenetic_2014}.
Research in the domain of data center demand management, which often utilizes load-shifting techniques, does not consider the caused carbon emissions but only addresses grid stability and energy prices \cite{Basmadjian_FlexibilityBasedEnergyDemandManagementCaseStudyCloud_2019, Liu_DemandResponseCoincidentPeak_2013, Klingert_SpinningGoldFromStraw_2020, Cioara_ExploitingFlexibilitySmartCitiesBusinessScenarios_2019}.
The practicability of temporal load-shifting with the goal to consume cleaner energy from the public power grid has only recently been demonstrated by Google's Carbon-Intelligent Computing System~\cite{Radovanovic_Google_2021} (CICS).
However, there does not yet exist any publicly available insights on the potential and theoretical limitations of this approach.

\vfill\break
\noindent
Addressing this gap, we make the following \textbf{contributions}:

\begin{itemize}
	\item we identify and categorize different characteristics of time-shiftable workloads in data centers.
	\item we define a methodology for estimating the regional carbon intensity of the public power grid using electricity production and inter-regional power flow data. 
	\item we analyze the carbon-saving potential of temporal workload shifting in four regions, namely Germany, Great Britain, France, and California.
	\item we experimentally evaluate two scenarios via simulation, examining the impact of time constraints, scheduling strategies, and the accuracy of forecasts.
	\item we make all data sets and code used for the analysis and experiments of this paper publicly available\footnote{Github: \url{https://github.com/dos-group/lets-wait-awhile}}.
\end{itemize}

The remainder of this paper is structured as follows. \autoref{sec:shiftable_workloads} discusses different characteristics of time-shiftable workloads. \autoref{sec:methodology} explains our methodology for the following analysis and evaluation. \autoref{sec:analysis} analyzes the theoretical potential for temporal workload shifting in four different regions. \autoref{sec:evaluation} experimentally evaluates two selected workload shifting scenarios via simulation. \autoref{sec:rw} reviews the related work. \autoref{sec:conclusion} concludes the paper.

\section{Shiftable Workloads}\label{sec:shiftable_workloads}
The most important properties for determining a workload's shifting potential are its time constraints.
While many workloads are expected to be finished as soon as possible, others may be subject to a degree of flexibility.
However, there are further properties, such as duration, execution time, and interruptibility, can have a substantial impact on whether and how a workload can be shifted in time.
This section categorizes workloads based on these characteristics.
The characteristics are experimentally evaluated regarding their impact in~\autoref{sec:evaluation}.
The terms workload and job are used interchangeably in this and the following sections.

\subsection{Duration}
While there is no consistent terminology, analyses of large cluster traces~\cite{Reiss_GoogleAnalysis_2012,Tirmazi_BorgNextGeneration_2020,Lu_AlibabaImbalance_2017,Guo_AlibabaEfficiency_2019} broadly categorize workloads into \textit{short-running}, \textit{long-running}, and \textit{continuously running}.

\subsubsection{Short-Running Workloads} 
Workloads executed in data centers are predominantly short-running.
An analysis has shown, that the majority of jobs in the Google cluster traces of 2011 last only a few minutes~\cite{Reiss_GoogleAnalysis_2012}. Similar findings were made on Alibaba cluster traces, where more than \SI{90}{\percent} of batch jobs run less than 15 minutes \cite{Lu_AlibabaImbalance_2017}, and are more likely to be deferred or evicted due to low priority levels~\cite{Guo_AlibabaEfficiency_2019}.
The shifting potential of such workloads highly depends on their time constraints. Most short-running workloads, such as Function-as-a-Service (FaaS) executions~\cite{Shahrad_CharacterizingServerless_2020} or CI/CD runs~\cite{Hilton_CIUsage_2016}, are expected to be finished in a timely manner. Even when delays of a few hours are tolerable, the expected potential for shifting is comparably small, as carbon intensity usually does not change quickly in large electrical grids. However, some batch jobs, such as nightly backups, may be accompanied by service-level agreements (SLAs) that allow for greater flexibility regarding the execution time. In these cases, the relative shifting potential is very high since the entire job can be moved to times of lower carbon intensity, and not only parts of it.

\subsubsection{Long-Running Workloads}
Analyses of Google cluster traces reveal that while only \SI{7}{\percent} of all workloads run at production priority, a majority of these jobs are long-running~\cite{Reiss_GoogleAnalysis_2012}. Thus, the resource and memory consumption of all jobs entail a heavy-tailed distribution, where a small portion of jobs consumes most of the resources~\cite{Reiss_GoogleAnalysis_2012,Tirmazi_BorgNextGeneration_2020}. Moreover, as shown on Alibaba cluster traces, long-running and prioritized workloads are likely to request significantly more resources and memory than they actually utilize~\cite{Lu_AlibabaImbalance_2017}.
For our paper, we define long-running workloads to have runtimes of up to several days. General examples for such jobs are machine learning trainings, scientific simulations, or big data analysis jobs. 
These workloads bear a notable absolute shifting potential since they are often very energy-intensive.
Moreover, it is often humans that rely on their results to take further action.
So, in practice, in many cases it makes no difference whether the issued job is finished in the middle of the night or the following morning.
This flexibility can be exploited by shifting workloads without interfering with the user's workflow.

\subsubsection{Continuously Running Workloads}
Many computational workloads, like user-facing APIs, effectively run indefinitely by design and cannot be interrupted. Apart from these so called continuous services, there exist other computationally intensive workloads such as blockchain mining, protein folding, brute force attacks, or very long-running scientific simulations, that execute over weeks and months, or do not have any defined end date. As an example, 2000 jobs of the Google cluster traces from 2011 run for the entire trace period of 30 days~\cite{Reiss_GoogleAnalysis_2012}.

Although blockchain mining in particular has received great attention for its immense power consumption~\cite{Krause_QuantificationEnergyCarbonCrypto_2018,Li_EnergyConsumptionMining_2019}, we do not consider these workloads as shiftable in this paper, as they have no deadline or a deadline very far in the future.
This paper only covers workloads up to several days, as real carbon intensity forecasts are based on weather and electricity demand forecasts which also only extend a few days into the future \cite{NationalGridESO_ForecastMethodology_2021, Lowry_DayAheadForecastingCarbonIntensityHVAC_2018, Bokde_ShortTermCO2ForecastingElectricityMarketScheduling_2021}.

\subsection{Execution Time}\label{sec:execution_time}
\begin{figure}[h]
    \centering
    \hspace{-0.3cm}\includegraphics[width=0.95\linewidth, trim=0cm 0.2cm 0cm 0cm, clip]{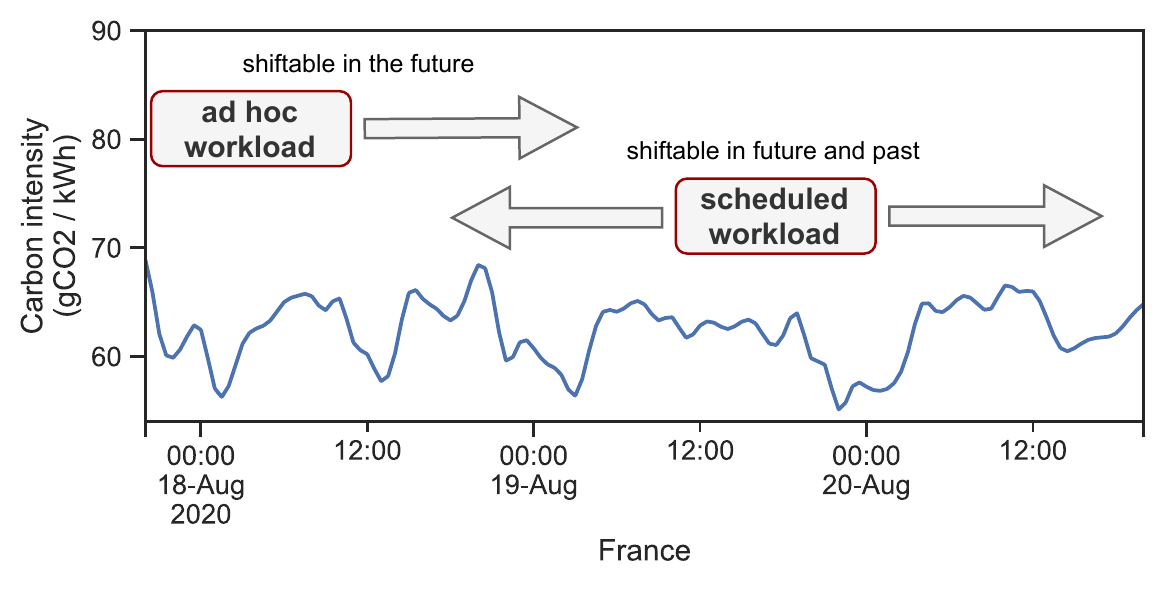}
    \vspace{-0.2cm}
	\caption{Scheduled workloads can potentially be shifted in both directions of time, while ad hoc workloads can only be deferred into the future.}
    	\label{fig:fig2}
    \vspace{-0.2cm}\end{figure}

The expected execution time of a workload and how strictly it should be enforced are important aspects in determining its shifting potential. We therefore elaborate two categories of execution time that are illustrated in \autoref{fig:fig2}.

\subsubsection{Ad Hoc Workloads}
A large number of workloads, short- and long-running, are issued in an ad hoc manner. Although some of them might follow a certain distribution which can be estimated by time series forecasting, it is not known upfront when exactly a specific job will be issued. Examples are again FaaS executions, CI/CD runs, machine learning trainings, and other jobs triggered by external events or issued by users for direct execution. 
The shifting potential of such workloads is limited to the future. In other words, only once a job is issued the scheduler can decide whether to execute the job immediately, or to postpone it under consideration of its time constrains.

\subsubsection{Scheduled Workloads}
We define scheduled workloads to be workloads that are planned to execute at a future point in time.
Prominent examples are periodically scheduled batch jobs such as nightly integration test suits, nightly builds, periodic backups, updates of search indices in databases, and auto-generated reports.
According to related work, a large number of jobs are recurring at fixed intervals.
For example, when comparing the Google cluster traces from 2011 to the traces of 2019, it can be observed that the  workload mix changed towards scheduled batch jobs while the scheduling rate increased significantly \cite{Tirmazi_BorgNextGeneration_2020}.
At Microsoft, periodic batch jobs have been reported to make up \SI{60}{\percent} of processing on large clusters~\cite{Jyothi_Morpheus_2016}. More than \SI{40}{\percent} of these jobs run on a daily basis, while other frequently used periods are fifteen minutes, an hour, and twelve hours. Another study revealed that recurring jobs make up roughly \SI{40}{\percent} of the jobs as well as cluster hours on all production clusters used for Microsoft’s Bing service~\cite{Agarwal_ReoptimizingDPC_2012}. 

Scheduled workloads can, depending on their time constraints, be shifted in both directions in time. For example, a nightly job which is usually executed periodically at 1~am, could also be scheduled at a more flexible time window between 23~pm and 3~am.

\newcolumntype{A}{>{\centering}m{0.075\textwidth}}
\newcolumntype{X}{>{\centering\arraybackslash}m{0.075\textwidth}}
\begin{table*}[ht!]
  \centering
  \small
  \caption{Carbon intensity of different energy sources according to \cite{IPPC_Annex2_2011}.}
  \begin{tabular}{|l||A|A|A|A|A|A|A|A|X|}
    \hline
    Energy Source & Biopower & Solar Energy & Geothermal Energy & Hydropower & Wind Energy & Nuclear Energy & Natural Gas & Oil & Coal \\
    \hline
    \coopkwh & 18 & 46 & 45 & 4 & 12 & 16 & 469 & 840 & 1001 \\
    \hline
  \end{tabular}
  \label{table:co2values}
\end{table*}

\subsection{Interruptibility}

\begin{figure}[b]
    \centering
    \hspace{-0.3cm}\includegraphics[width=0.95\linewidth, trim=0cm 0.2cm 0cm 0cm, clip]{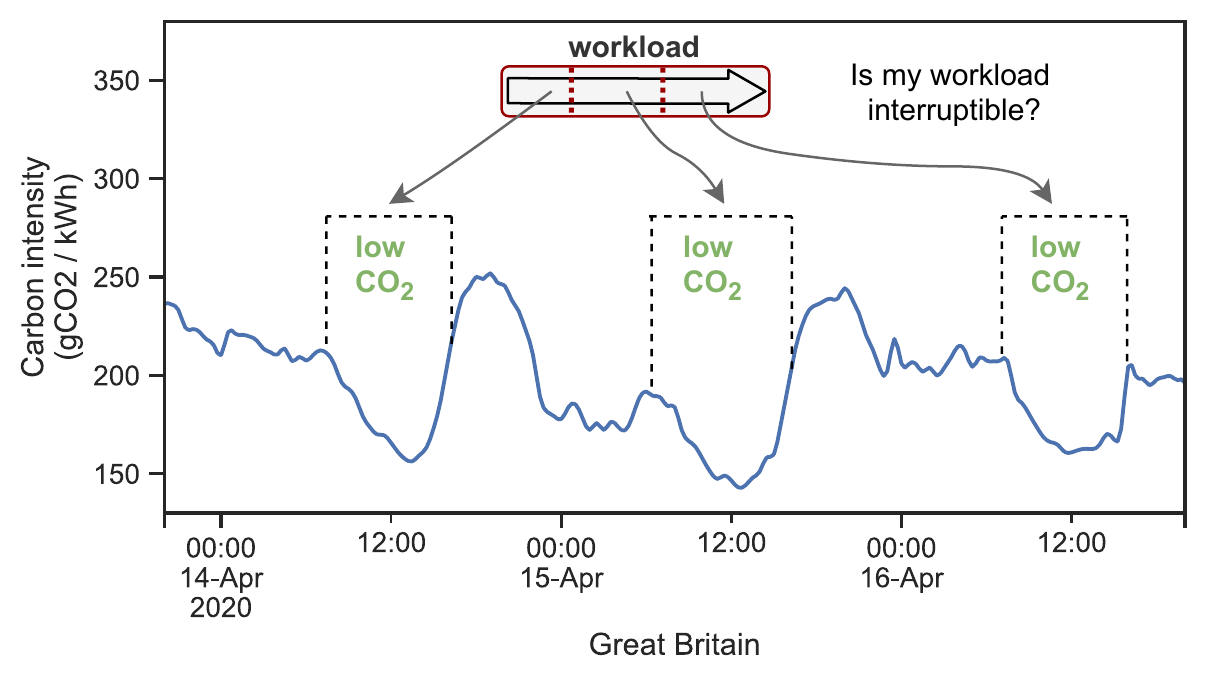}
    \vspace{-0.2cm}
	\caption{Interruptible workloads can be divided into chunks and scheduled separately.}
    	\label{fig:fig3}
    \vspace{-0.2cm}
\end{figure}

While certain workloads incorporate checkpoint mechanisms or store intermediate results and can thus be paused and resumed at a later point in time, other workloads must be executed without interruption.
As the interruptibility of workloads can be exploited by carbon-aware schedulers as depicted in \autoref{fig:fig3}, to better align the load to times of low-carbon energy, we categorize workloads according to their interruptibility.

\subsubsection{Interruptible Workloads}
The possibility of pausing and resuming jobs is frequently seen in long-running workloads.
Prominent examples are iterative machine learning trainings or discrete-event simulations, which often periodically write checkpoints for later analyses, resumption from earlier states, and error handling.
By using such checkpoint mechanisms and state handling, it is possible to interrupt and resume workloads at a later point in time~\cite{Ferreira_CheckpointingExtremeScaleComputing_2014,Rojas_CheckpointingDNNs_2020}.
Further examples of interruptible workloads include jobs that consist of many smaller tasks, like the generation of monthly business reports for different clients.
As the carbon intensity of large, interconnected regions does usually not change with high frequency, it is not meaningful to split workloads in very small chunks.
From this follows that the overhead, which arises when stopping and starting jobs, can often be neglected.

\subsubsection{Non-Interruptible Workloads}
Other workloads cannot be interrupted or interrupting them is not practical because the energy cost of starting and stopping the work outweighs the expected benefit.
Examples include certain CI/CD or compile jobs that often run in freshly created, encapsulated environments which need a significant amount of time for setup and tear-down.
Database migrations and backups are usually required to execute in one go to avoid data inconsistencies.
Additionally, many test suits cannot be interrupted by design, for example, when they test a system under load.
Non-interruptible workloads always have to be scheduled in one consecutive period and are, hence, less flexible when it comes to avoiding local maxima in carbon intensity.

\section{Regional Carbon Intensity}\label{sec:methodology}

This section describes our methodology for selecting the analyzed regions, collecting data, and calculating the average carbon intensity of regions over time. 
As we want to publish all used datasets, we did not use commercially available data such as offered by services like \emph{electricityMap}\footnote{\url{electricitymap.org}, accessed 2021-09-21}.
All following analyses and experiments base on the data described in this section.

\subsection{Region Selection}\label{sec:region-selection}

Our analysis covers four different regions: Germany, Great Britain, France, and California. Regions were selected by the following three criteria:

\begin{enumerate}
  \item[(1)] \emph{Representativeness}: To represent relevant locations for data center operation, we only chose regions in which the three biggest public cloud providers - AWS, Microsoft Azure, and Google Cloud - offer regions or availability zones, or have publicly announced plans to launch operations in the near future.
  \item[(2)] \emph{Availability of data}: For our analysis we require access to each region's electricity production data by energy source with at least hourly reporting interval for the entire year 2020. 
  \item[(3)] \emph{Regional diversity}: Selected regions should have different characteristics regarding types and extent of utilized energy sources as well as geographic location, to represent a diverse spectrum of regional differences.
\end{enumerate}

Unfortunately, the second criteria eliminates many candidate regions because the availability of open access data on electricity production by energy source is limited. We would have liked to include regions from the southern hemisphere and emerging markets such as Brasil, South Africa, India, Korea, Japan, or Australia. However, for none of these regions it is currently possible to access historical data in the quality required for this study. Consequently, Criteria (3) is only fulfilled partially: While our selected regions do have diverse characteristics, all are located in Europe or the US.

\subsection{Carbon Intensity of Energy Sources}

The carbon intensity (\coopkwh) of an energy source describes the amount of carbon emitted per \si{kWh} of electricity produced. 
There exist numerous studies on the carbon intensity of different energy sources that use slightly varying methodologies and base their estimates on different data. We base our research on carbon intensity estimates that take into account the whole life-cycle of energy sources. 
In particular we use the data from a comprehensive IPCC literature review that determined the median carbon intensity value stated by hundreds of different studies~\cite{IPPC_Annex2_2011}. The values are presented in \autoref{table:co2values}.

\begin{figure}[b]
    \centering
        \includegraphics[width=0.9\linewidth]{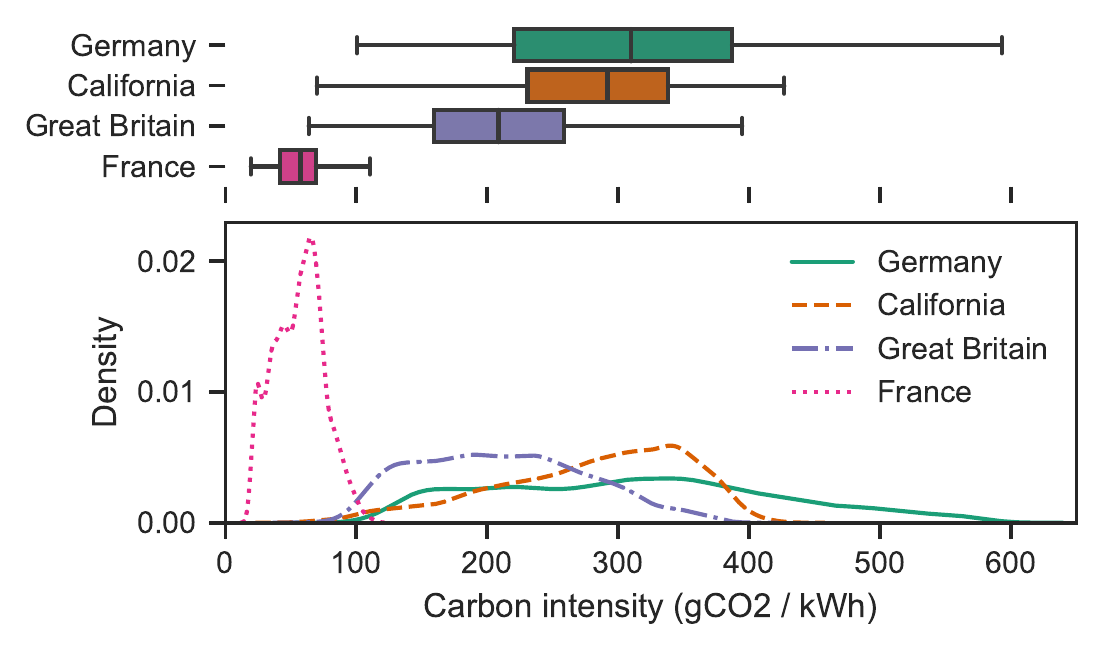}
		\caption{Distribution of carbon intensity values in the four observed regions in 2020.}
    	\label{fig:hist1}
\end{figure}

\begin{figure*}[t]
    \centering
    \includegraphics[width=1\linewidth]{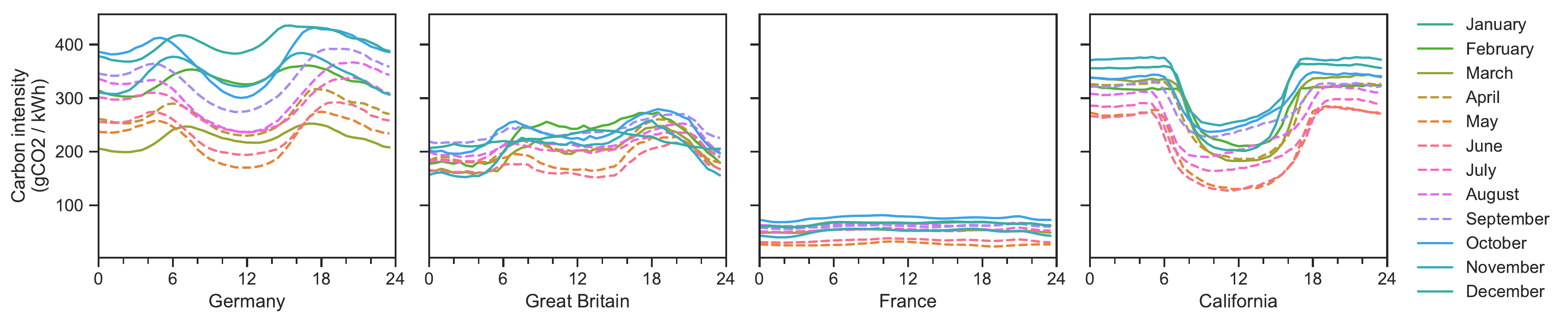}
	\caption{Daily mean carbon intensity of Germany, Great Britain, France and California by month. Since all regions are located in the northern hemisphere and therefore exhibit similar seasonal patterns, we use the same cyclic colormap to illustrate the differences between winter (solid lines) and summer months (dashed lines).}
    	\label{fig:seasons}
\end{figure*}

\subsection{Carbon Intensity of Regions}

To better represent the carbon emissions data centers cause by \emph{consuming} energy, additionally to regional energy production we also consider cross-regional flows of energy. 
The most precise method for this so called consumption-based accounting is to calculate the carbon intensity of all neighboring regions and to apply flow tracing in order to resemble the underlying physics of the power grid~\cite{Tranberg_CarbonAccounting_2019}.
Since detailed energy generation data is not available for many regions and cross-border flows usually do not amount to a large fraction of the available power, we use a simplified method and only consider the yearly average of the neighboring regions to weight their contribution.

We define the average carbon intensity of a region $C_t$ at time~$t$ by weighting the power generation $P_{g,t}$ of each energy source $s \in S$ by its respective carbon intensity~$c_s$. As explained above, we additionally weight each energy import from neighboring regions $r \in R$ by the average carbon intensity of that region~$c_r$. The resulting sum is divided by the sum of all generated and imported electricity:
$$C_t = \frac{\sum\limits^S_{s=1} P_{s,t}\ c_s + \sum\limits^R_{r=1} P_{r,t}\ c_r}{\sum\limits^S_{s=1} P_{s,t} + \sum\limits^R_{r=1} P_{r,t}}$$

For our analysis we consider the entire year 2020. The electricity production and cross-border flow data for all three European regions were retrieved via the ENTSO\nobreakdash-E (European Network of Transmission System Operators for Electricity) API\footnote{\url{https://transparency.entsoe.eu}, accessed 2021-09-21}. Data from the California region were retrieved via CAISO (California Independent System Operator)\footnote{\url{http://www.caiso.com}, accessed 2021-09-21}. All data were adjusted to a common resolution of 30 minutes.
For electricity production, we mapped the returned energy sources to the categories stated in \autoref{table:co2values}. 
For cross-border flows, we used the yearly average carbon intensity of neighboring regions for 2020 \cite{CountryCarbonFootprint_2020}.

\subsection{Average vs. Marginal Carbon Intensity}\label{sec:average_vs_marginal}

Our methodology calculates the \emph{average} carbon intensity of regions, namely their current electricity mix weighted by the carbon intensity of energy sources.
A signal that captures the cause-effect relationship of load shifting even better is the \emph{marginal} carbon intensity, which describes the carbon emissions of the energy source responsible for generating additionally requested electricity at given point in time.

Unfortunately, in practice it is very hard to identify this marginal energy source, as the decision of a power supplier to scale their production up or down is not centralized but usually incentivized via electricity prices.
Additionally, this decision depends on a variety of further factors such as forecasted weather and demand as well as expected surplus or demand in neighboring regions.
For this reason, there exist only probability-based methods to compute marginal carbon intensity whose results fluctuate depending on the region and time of day~\cite{Leerbeck_TomorrowCo2Forecasting_2020}.
After reviewing marginal data provided by electricityMap, we consider marginal carbon intensity to be no practical signal for demand management at this point due to high uncertainties.
This assumption is supported by Google's CICS, that also uses the average carbon intensity as an indicator for their load shifting efforts.

\section{Analysis of Theoretical Potential}\label{sec:analysis}

We examine the energy mix and resulting carbon intensity over time in Germany, Great Britain, France, and California throughout the year 2020. This section aims at identifying patterns in this data that can be exploited by temporal workloads shifting.

\subsection{Region Analysis}
In the following, the properties and peculiarities of the energy mix in the four selected regions as well as the statistical moments of their resulting carbon intensity are described.
The distribution of carbon intensity values is displayed in~\autoref{fig:hist1}. The average carbon intensity throughout a day is presented in~\autoref{fig:seasons} for each month and region.

\subsubsection{Germany}

Due to the wide adoption of wind (\SI{24.7}{\percent}) and solar power (\SI{8.3}{\percent}), one third of the German electricity production comes from highly variable, renewable sources. On the other hand, the remaining electricity mix is disproportionately dirty, as it is largely generated by burning lignite and black coal (\SI{22.8}{\percent}) as well as fossil gas (\SI{11.3}{\percent}). This discrepancy translates into both, the highest mean carbon intensity of \SI{311.4}{\coopkwh} across all observed regions, as well as highest variation of values, reaching from 100.7 to \SI{593.1}{\coopkwh}. The mean daily carbon intensity varies greatly over the year with a difference of up to \SI{100}{\percent}. However, the inner-daily variance remains approximately equal regardless of the season. We observe that energy is usually the cleanest during mid day, when most solar energy is available, and around 2 am, when electricity demand is generally low and fossil fuel power plants are throttled back.

\subsubsection{Great Britain}

Great Britain relies mainly on burning fossil gas (\SI{37.4}{\percent}), wind power (\SI{20.6}{\percent}) and nuclear energy (\SI{18.4}{\percent}). It has a comparably diverse energy mix and only roughly \SI{8.7}{\percent} of the consumed energy is imported. The mean carbon intensity of \SI{211.9}{\coopkwh} and standard deviation are considerably lower than in Germany and stays approximately equal over the year. The inner-daily variance is higher in the winter months. Like in Germany, the carbon intensity is the cleanest during night time. However, due to the lower deployment of solar energy, carbon intensity does not drop as significantly during daylight hours.

\subsubsection{France}

The French energy mix comprises \SI{69.0}{\percent} of nuclear power and \SI{8.6}{\percent} of hydropower. Both of these energy sources are characterized by very low carbon emissions and low variability. Only a little more than \SI{10}{\percent} of the electricity stems from variable renewable sources like wind and sun. As a result, the French power grid's carbon intensity is not only very low throughout the entire year, with a mean of \SI{56.3}{\coopkwh}, but also very steady. Likewise, the inner-daily variance is comparably low.

\subsubsection{California}

California generates \SI{13.4}{\percent} of its total electricity from solar power - in the period between 8\,am and 4\,pm even \SI{30.9}{\percent}. On the other hand, one third of the electricity comes from fossil gas and more than one quarter of the energy is imported from neighbouring states that have a comparably dirty energy mix. As a result the mean carbon intensity of \SI{279.7}{\coopkwh} is almost as high as in Germany, although the range of values is more comparable to Great Britain. Nevertheless, \autoref{fig:seasons} shows that California has very different characteristics than these regions. Because of the large amount of solar energy, the length of the low carbon intensity window during the day is strongly correlated with the number of hours of sunshine in a given month. The mean carbon intensity is generally lower in the summer months than in the winter months.

\subsection{Weekly Patterns}

\begin{figure}[h]
    \centering
    \includegraphics[width=0.9\linewidth]{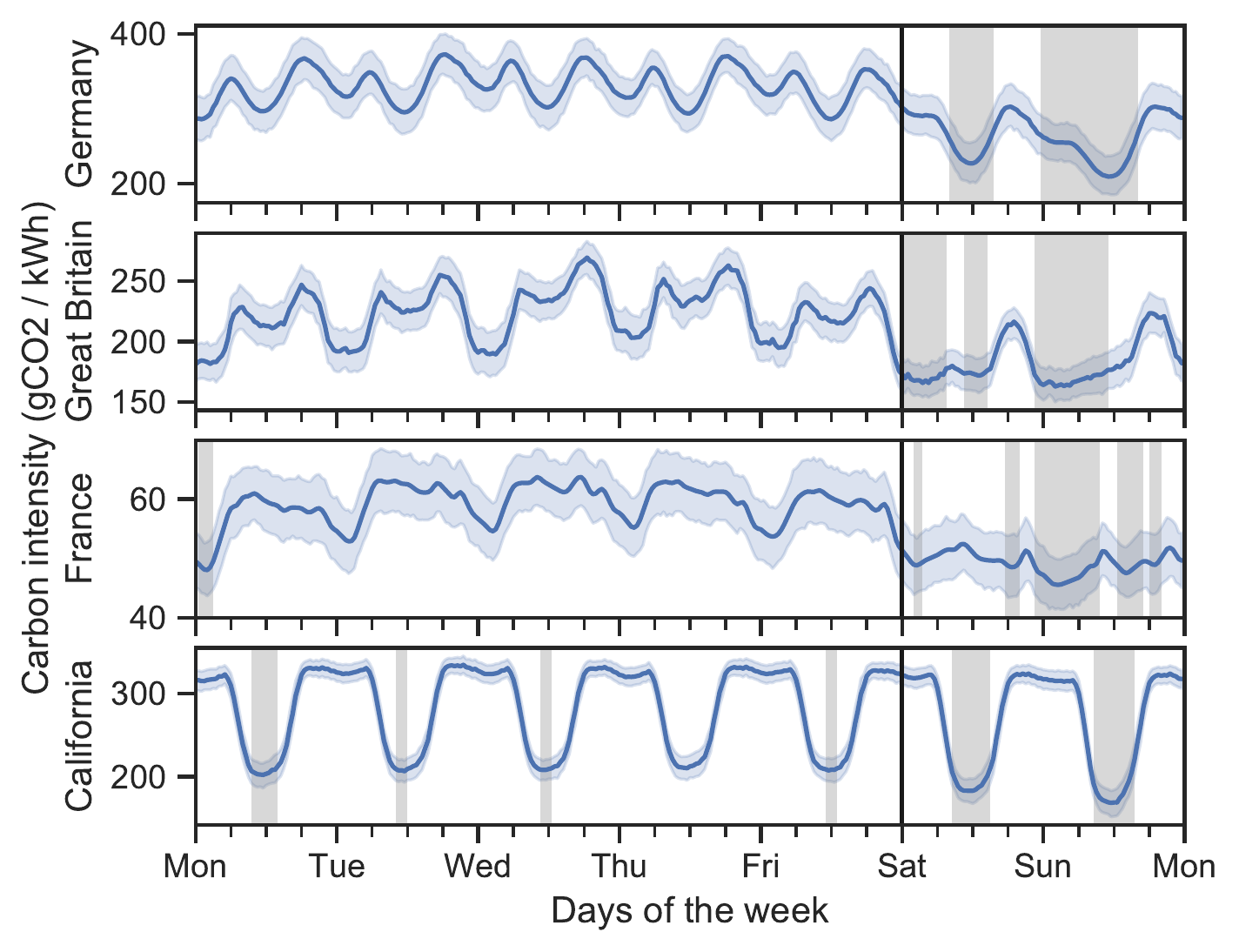}
	\caption{Mean carbon intensity during a week. The confidence interval describes the 95th percentile. Highlighted in gray are the 24 hours with lowest carbon intensity, which predominantly fall on the weekend across all regions.}
    	\label{fig:week}
\end{figure}

As some non-urgent workloads can be postponed by multiple days, we first observe weekly seasonal patterns that can be exploited as shown in \autoref{fig:week}.
We observe that the daily carbon intensity behaves similar during workdays but has a clear drop during weekends. For example, carbon intensity of an average workday in Germany is \SI{328.7}{\coopkwh}, the average value during weekends is only \SI{243.7}{\coopkwh}, which is a decrease of \SI{25.9}{\percent}. Likewise, we can observe decreased carbon intensity on weekends in Great Britain (\SI{20.7}{\percent}), France (\SI{22.2}{\percent}), and California (\SI{6.2}{\percent}). 

This drop is caused by the decreased power demand on weekends which electricity providers respond to by mainly reducing the amount of power produced by fossil fuels. For instance, on average Germany produces \SI{28.7}{TW} of energy on workdays and only \SI{21.2}{TW} on weekends.
The fact that electricity is greener on weekends across all observed regions suggests that shifting load to weekends is a promising approach in general. However, as stated above, it is limited to workloads with relaxed time constraints. 

\subsection{Best Times of Day for Shifting}

To identify the most promising times of day for shifting workloads, we define the shifting potential $p(t, W)$ at time $t$ as follows:
$$p(t, W) = C_t - \min_{\forall t^\prime \in W} C_{t^\prime}$$

\begin{figure*}
    \centering
    \includegraphics[width=1\linewidth]{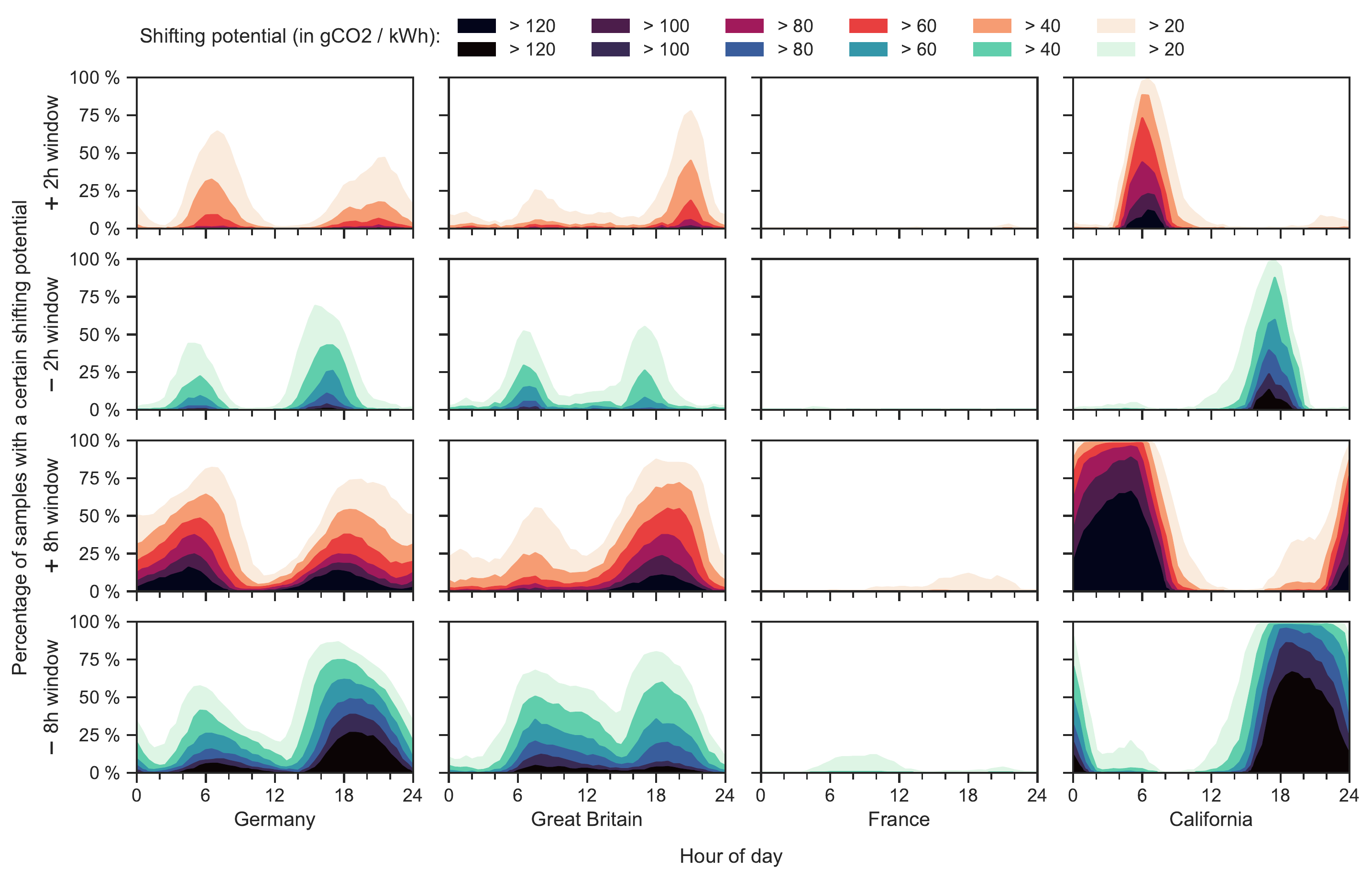}
	\caption{Potential for shifting workloads in the future (+) or past (-) at different times of the day and at two exemplary flexibility windows; 2 hours and 8 hours. For instance, the plot in column 1, row 3 describes the potential of shifting workloads in Germany by up to 8 hours into the future: The carbon intensity of \SI{14}{\percent} of the workloads scheduled at 5\,am could be reduced by at least \SI{120}{\coopkwh} when instead scheduled between 5\,am and 1 pm.}
    	\label{fig:potential}
\end{figure*}

\noindent
where $W$ describes the forecast window, namely the set of carbon intensity data points following or preceding $t$. Intuitively, this function describes by how much the carbon intensity could theoretically be reduced when shifting a short-running workload at time $t$ for up to $W$ into the future or past. Shifting into the "past" is of course only possible for workloads that are scheduled for future execution (see \autoref{sec:execution_time}). The carbon intensity of regions does usually not change rapidly, nor is the signal very noisy. This is why searching for the minimum value is a suitable metric here, as the chance of optimizing for negative spikes in the signal noise is very low. The presented metric only considers single data points, in other words workloads of up to 30 minutes of length, and assumes we have perfect forecast accuracy.

\autoref{fig:potential} displays the shifting potential of all regions aggregated by the time of day throughout the year for four different windows: Shifting into the future and past by a maximum of two or eight hours.
When considering the first row, namely shifts of up to two hours in the future, most regions exhibit little potential. An exception is California where there is a considerable shifting potential before sunrise, when carbon intensity usually drops heavily. For example, at \SI{44}{\percent} of the days in 2020, the carbon intensity of workloads scheduled at 6\,am could be reduced by more than \SI{80}{\coopkwh} if instead scheduled between 6 and 8\,am. Scheduled workloads can also be shifted in the opposite direction, as presented in the second row. Again, California shows the highest potential by shifting load from after to before sunset.

It becomes apparent that with bigger forecast window size the potential for improvement increases substantially. However, the optimal times for shifting differs highly in the observed regions.
In Germany, we observe two times of the day that show potential for load shifting at 8 hour windows: In the morning hours around 7\,am before sunrise and around 6\,pm, escaping the high-carbon evening hours.
Nevertheless, due to the high variability of energy sources in Germany, such larger forecasts offer a certain potential at virtually any time of day.

The potential for shifting workloads into the future during morning hours is considerably smaller in Great Britain, but comparably big in the evening. In general, we can observe that there is almost no potential in both directions during night time.
As expected, there is barely any load shifting potential in France, even at large forecast windows. This is due to the already low carbon intensity and low variability of values during a day.
In California, the potential for large forecast windows is very high during nighttime, due to the steep drop in carbon intensity during daylight hours. Consequently, workloads that are already scheduled during daytime, show little to no potential.

The key finding from this analysis is that the potential for load shifting into the future, which can be exploited by all shiftable workloads, is generally highest in the early morning hours for countries with a lot of solar power and in the evening hours for countries that throttle their fossil fuel production at night.
Load shifting into the "past", which can only be exploited by future scheduled workloads, holds just as much potential and can in most cases complement load shifting into the future to attain potential throughout most parts of the day.

\section{Experimental Evaluation}\label{sec:evaluation}
So far, we have analyzed the theoretical potential of temporal workload shifting. In this section, we evaluate two realistic load shifting scenarios experimentally, examining the effects of time constraints, scheduling strategies, and forecast errors. Since openly available cloud computing data sets that contain information about the delay-tolerance of workloads are not available, we created two scenarios ourselves, featuring (1) short-running, periodically scheduled jobs, and (2) long-running machine learning trainings based on the StyleGAN2-ADA \cite{Karras_StyleGAN2ADA_2020} paper.
The experiments are simulated using LEAF \cite{Wiesner_LEAF_2021}, an IT infrastructure simulator that enables high-level modeling of energy consumption. The experimental setup comprises a single node, representing a data center, on which the jobs are scheduled.

\subsection{Scenario I: Nightly Jobs}
\label{sec:evaluation_periodic_jobs}

In the first scenario, we simulate a periodically scheduled job, such as a nightly build, integration test, or database migration. We assume these jobs to be delay-tolerant in most cases, meaning it does not make a difference to the user when exactly the job is executed, as long as it is outside of working and high-traffic hours. The aim of this experiment is to investigate the carbon saving effect of increasing the scheduling flexibility.

\subsubsection{Experimental Setup}\label{sec:evaluation_scenario1_setup}

We simulate 366 periodically scheduled jobs, one for each day of the entire year 2020, with a step size of 30 minutes. Likewise, each job takes 30 minutes and is not interruptible.
In the baseline experiments, jobs are scheduled to always run at 1~am.
For every region, we run 16 more experiments, each increasing the time window for scheduling jobs by 30 more minutes in both directions.
For example, the first shifting experiment executes all jobs between 12:30 and 1:30~am, the second between 12 and 2~am, and the last experiment schedules jobs between 5~pm and 9~am. 

Since openly available, ready-to-use solutions for forecasting grid carbon intensity across different regions are not available (see \autoref{sec:rw_forecasts}), we added noise to the observed carbon intensity timeline in order to simulate inaccurate forecasting results. 
We calculated a mean absolute error of 10 for the 48-hour carbon intensity forecast by National Grid ESO \cite{NationalGridESO_ForecastMethodology_2021} for 2020, which is roughly \SI{5}{\percent} of its yearly mean.
Based on this, we ran all experiments by applying normally distributed noise with $\sigma=0.05 $ times the yearly mean of the regional carbon intensity. The noise is independent of the forecast length. Since predictions in this scenario are at most 16 hours, this error can be considered an upper limit. Additionally, we repeated all experiments with optimal forecasts to investigate on the impact of errors. All experiments with forecast errors were repeated ten times and averaged.

\begin{figure}[b]
    \centering
    \includegraphics[width=0.85\linewidth]{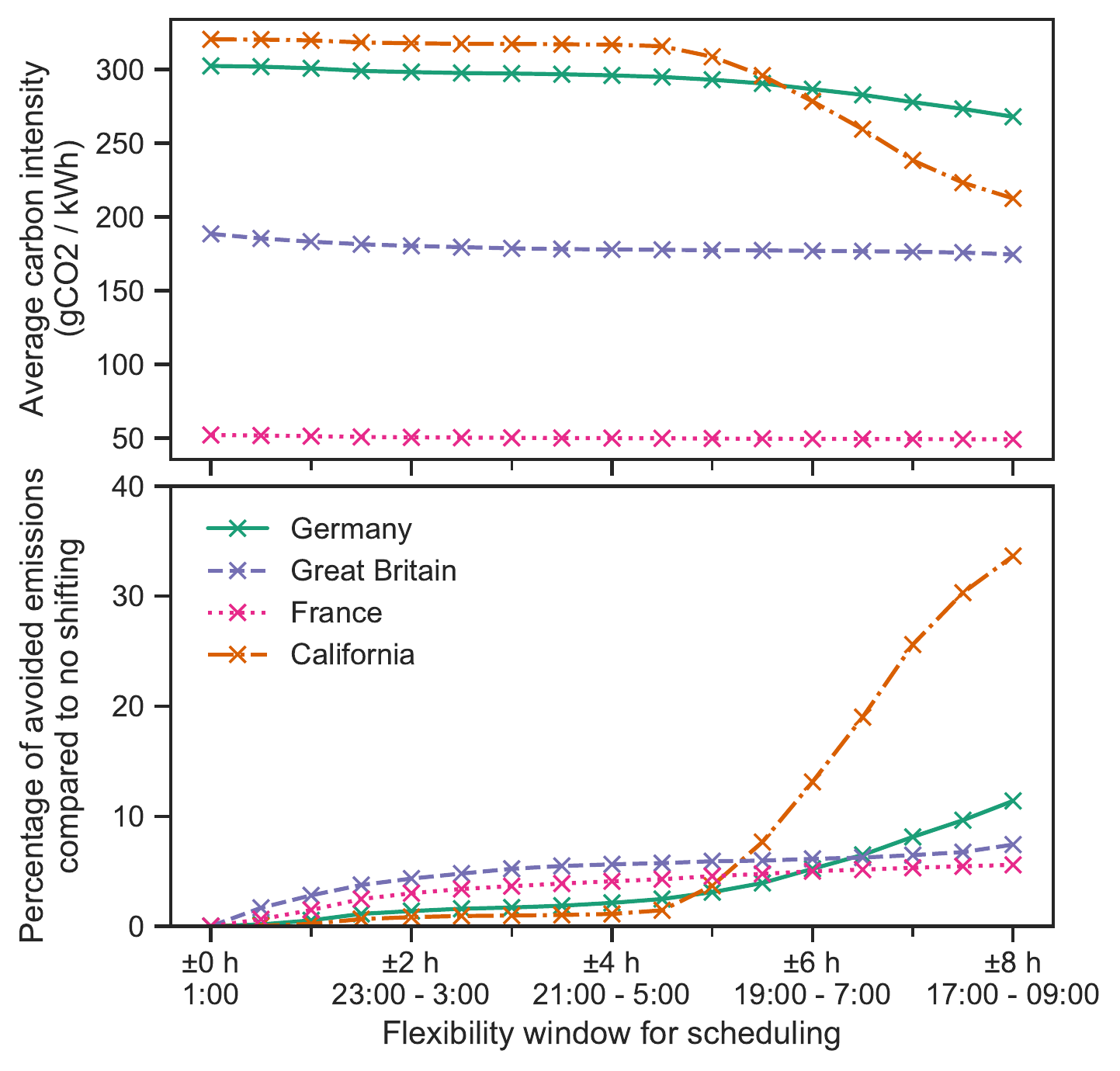}
	\caption{Average grid carbon intensity at job execution time. With increasing flexibility, the achieved carbon savings increase as well. The forecast error is \SI{5}{\percent} in all experiments.}
    	\label{fig:nightly_result}
\end{figure}

\subsubsection{Results}

\autoref{fig:nightly_result} displays the experimental results.
We can observe that relative to the baseline, carbon savings can be achieved across all regions. The effects differ significantly depending on the region and scheduling flexibility. 
For example, in France and Great Britain we can already achieve savings of \SI{3.0}{\percent} and \SI{4.3}{\percent}, respectively, when increasing the flexibility window by only $\pm 2$ hours. However, when the window is further enlarged, little additional savings are observed. For example, in France, the average grid carbon intensity used for powering the jobs could only be reduced by \SI{4.1}{\percent} when considering the $\pm 8$ hour window at \SI{5}{\percent} forecast error. 
In Great Britain, we managed to save \SI{7.4}{\percent} of carbon over the year with these parameters.

Ai flexibility windows of up to $\pm 4$ hours, the resulting emissions savings for Germany and California are almost negligible.
However, we observe a steep increase for windows starting at $\pm 5$ hours
Even when considering forecast errors, the German scenario emits \SI{11.2}{\percent} less carbon for the $\pm 8$ hour experiment.
The forecast error has a considerable impact on this result; carbon savings were more than 2 percentage points higher with optimal forecasting.
In California, the increased flexibility accounts for \SI{13.1}{\percent} savings for the $\pm 6$ hour window and \SI{33.7}{\percent} for the $\pm 8$ hour window under forecasts with error.
The impact of errors is less significant here; optimal forecasting only improves these results by 1-1.5 percentage points.

\subsubsection{Discussion}

\begin{figure}
    \centering
    \includegraphics[width=1\linewidth]{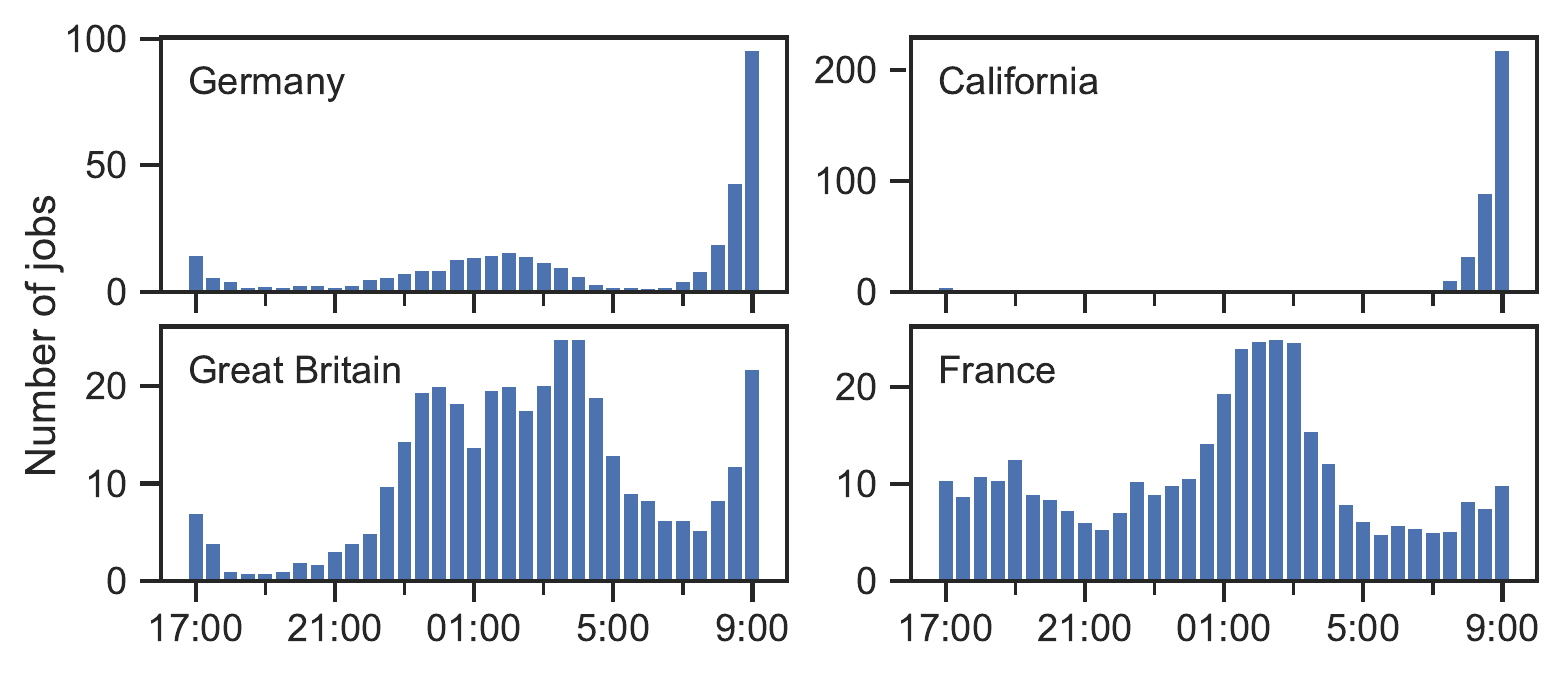}
	\caption{Number of jobs by allocated time slot for $\pm 8$ hour window size and \SI{5}{\percent} forecast error. Germany and California shift heavily into morning hours, while Great Britain and France distribute jobs more evenly during the night.}
    	\label{fig:nightly_hist}
\end{figure}

The results are consistent with our analysis on the shifting potential at different times of the day, see \autoref{fig:potential}. In France and Great Britain shifting potential is comparably low at night, because the mean carbon intensity at this time is already at its minimum.
In contrast, in Germany and California, the potential grows significantly once the scheduler has the ability to shift workloads to the early morning or late evening hours, where they can benefit from solar energy generated during the day.
This assumption is backed by \autoref{fig:nightly_hist}, which shows the number of jobs that were allocated to certain time slots in different regions.

For California, the case is simple:
Scheduling "nightly" jobs to after sunrise significantly reduces their carbon emissions. Also in the other regions carbon-aware scheduling can reduce emissions by more than \SI{10}{\percent}. This is not insignificant, given that the proposed shifting strategy does not have any negative impact on data center operations.
From a service provider perspective, these findings can influence the design of future service-level agreements (SLAs) and, hence, middleware systems that act within their boundraries.
Providing time windows instead of fixed points in time for service execution appear to be easy-to-implement measures for reducing the carbon footprint of cloud services.

\subsection{Scenario II: Machine Learning Project}

The second scenario investigates the impact of different workload shifting strategies on a large machine learning project comprising a variety of different jobs.
The scenario is based on the energy consumption statistics published for transparency reasons with a recent paper by NVIDIA Research introducing the StyleGAN2-ADA \cite{Karras_StyleGAN2ADA_2020} model. The paper has received attention not only for its novelty in training generative adversarial networks, but also because the authors required \SI{325}{MWh} of energy in the process of doing their research, suggesting large potential for carbon savings.

\subsubsection{Experimental Setup}

The authors of \cite{Karras_StyleGAN2ADA_2020} state that 3387 machine learning jobs were executed for creating the paper, worth $145.76$ GPU years. Their jobs usually run on eight GPUs, hence, an average job takes almost two days. In our scenario we assume that all jobs are scheduled ad hoc and randomly distributed across all 262 workdays of 2020 by sampling from a multinomial distribution.
Each jobs is assigned a random start time during core working hours (Monday to Friday, 9~am to 5~pm). Job durations are evenly distributed between four hours and four days, resulting the same amount of GPU years as in the original project. Furthermore, we assume that job durations are known upfront accurate to 30 minutes, which is the simulation step size.

Our baseline experiment starts all jobs right when they are issued.
We evaluate the potential of workload shifting in this scenario based on two time constraints:

\begin{description}
	\item[Next Workday] If jobs finish during non-working hours, they can be shifted as long as they finish before the next working day at 9~am. This allows the scheduler to take advantage of jobs that would otherwise be finished during the night or weekend without interfering with the workflow of researchers. In our scenario, this time constraint results in \SI{20.4}{\percent} of jobs that are not shiftable because they end during working hours, \SI{51.2}{\percent} are shiftable until the next morning and \SI{28.4}{\percent} are shiftable over the weekend.
	\item[Semi-Weekly] In practice, the individual results are often not required directly at 9~am the next day, but are evaluated in larger batches. If the time where results are actually required will be provided by users, the flexibility window for scheduling and, hence, saving potential can increase substantially. To represent this circumstance in a simple way, we assume in this time constraint that machine learning results are evaluated only twice a week. Concretely, all jobs can be shifted until the next Monday or Thursday at 9~am.
\end{description}

Furthermore, we want to investigate the potential benefits of exploiting incorruptible jobs, like machine learning trainings, by evaluating two scheduling strategies:

\begin{description}
	\item[Interrupting] The scheduler searches for the individual 30 minute intervals with the lowest carbon intensity and splits the job execution among these intervals.
	\item[Non-Interrupting] The scheduler searches for the coherent time window with the lowest average carbon intensity and does not split the job execution.
\end{description}

We simulate all combinations of time constraints and scheduling strategies for each country, with a \SI{5}{\percent} forecast error as described in \autoref{sec:evaluation_scenario1_setup}.

\begin{figure}[b]
    \centering
    \includegraphics[width=1\linewidth]{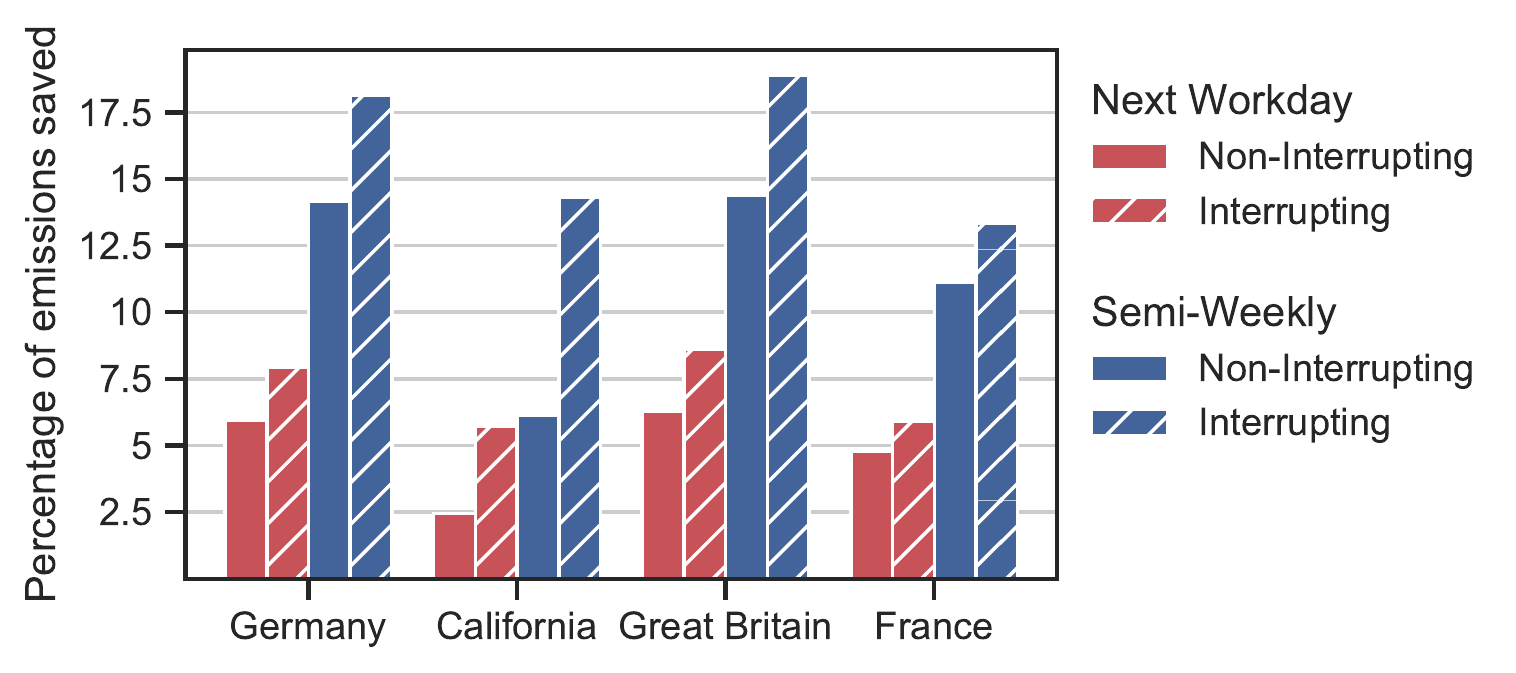}
	\caption{Carbon emission savings for different scheduling constraints and strategies by region. All experiments were simulated with 5\,\% forecast error.}
    	\label{fig:ml_results}
\end{figure}

\begin{figure}
    \centering
    \includegraphics[width=0.9\linewidth]{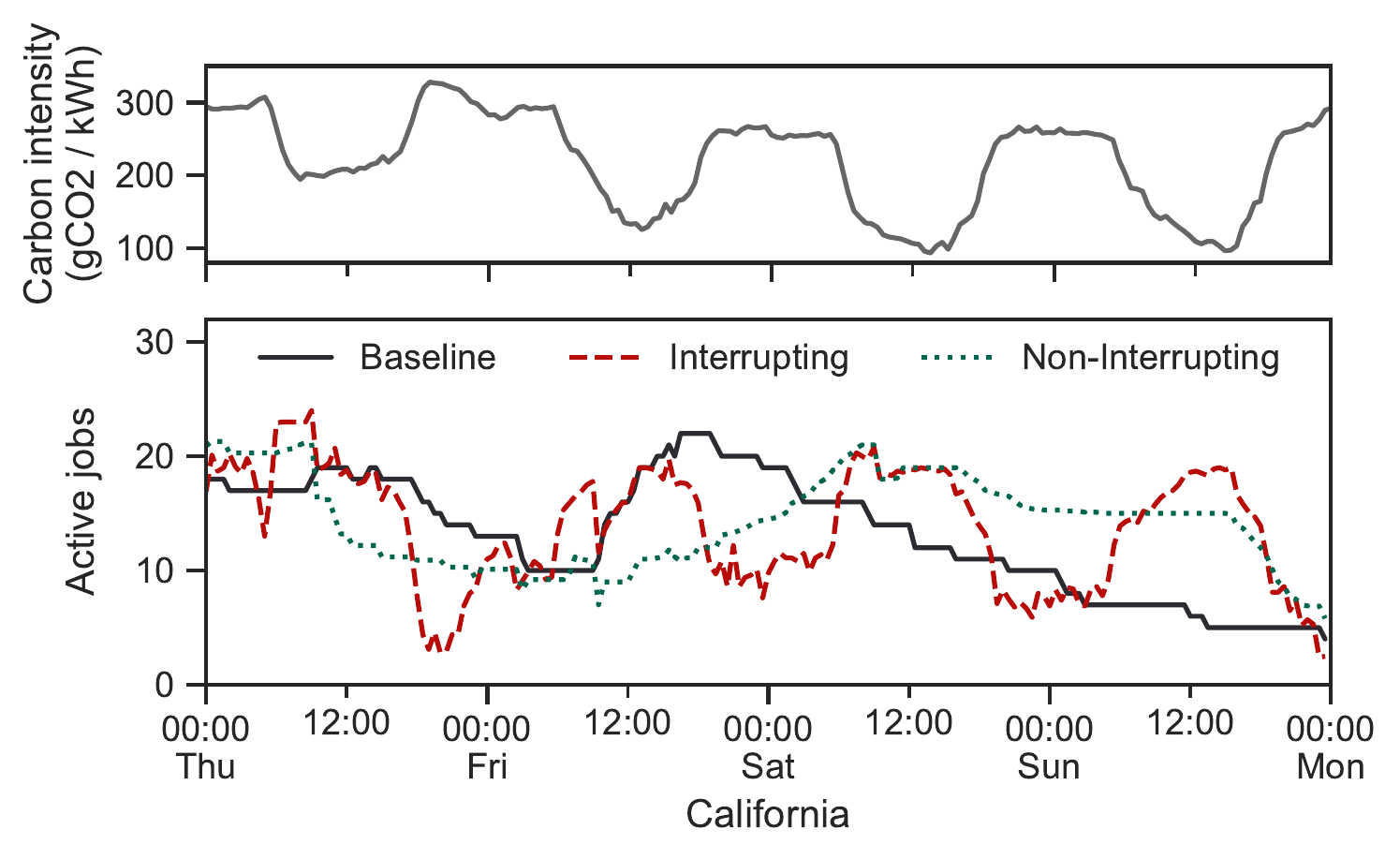}
	\caption{Number of active jobs over time for different scheduling strategies compared to the current carbon intensity. Data is from the California region, June 4-7.}
    	\label{fig:ml_jobs}
\end{figure}

\begin{figure}
    \centering
    \includegraphics[width=0.9\linewidth]{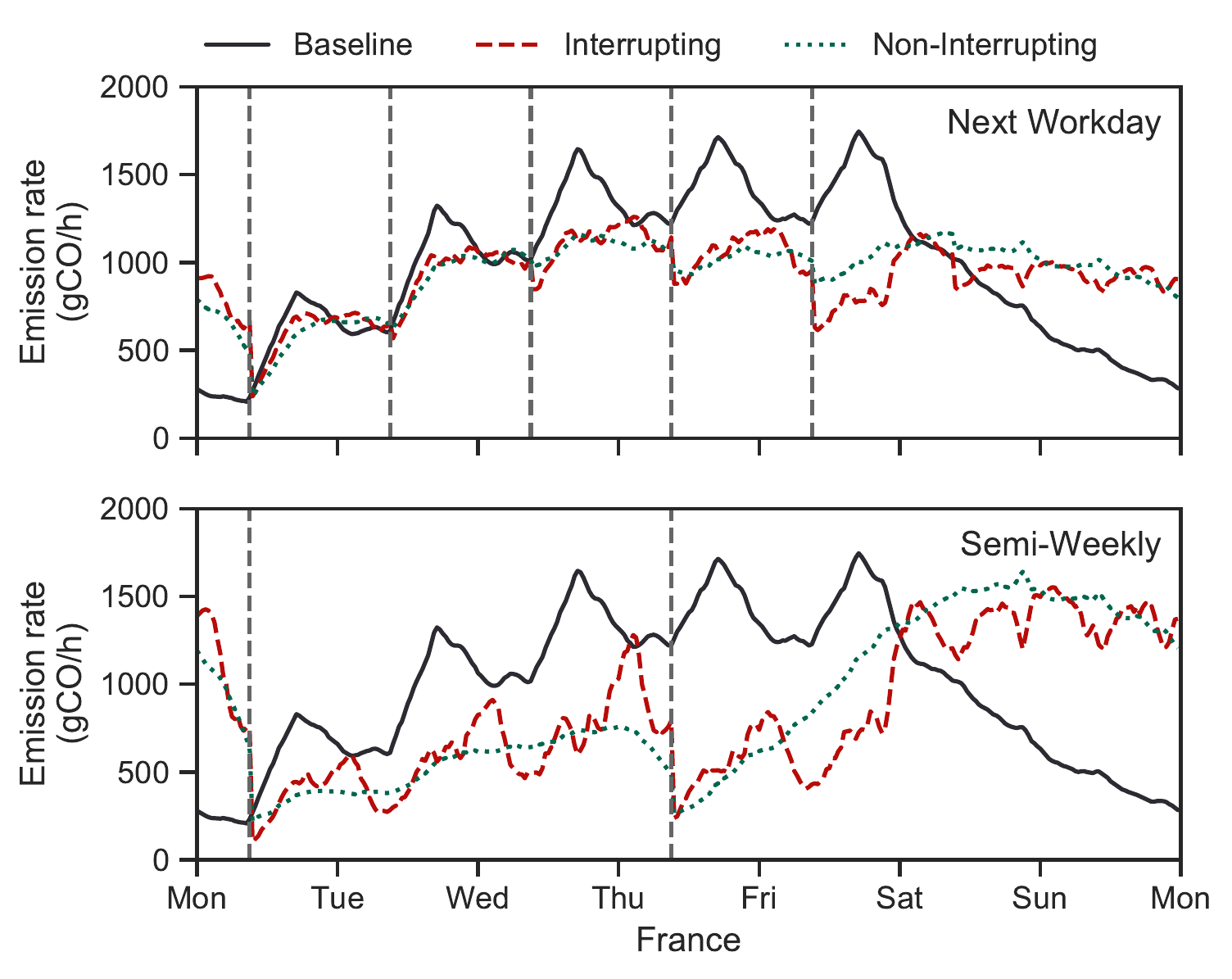}
	\caption{Average emission rates caused by different scheduling scenarios during an average week. Gray dashed lines represent the deadlines of when jobs are supposed to be finished at the two different time constraints.}
    	\label{fig:ml_constraints}
\end{figure}

\subsubsection{Results}\label{sec:ml_results}

The carbon savings achieved by the experiments relative to the respective region's baseline experiment are depicted in \autoref{fig:ml_results}. 
When considering the Next Workday constraint, the Non-Interrupting scheduling managed to reduced the project's carbon emissions by \SI{2.5}{\percent} to \SI{6.3}{\percent}, while the Interrupting scheduling achieved reductions of \SI{5.7}{\percent} to \SI{8.5}{\percent}.
For the Semi-Weekly constraint, Non-Interrupting scheduling saved \SI{6.1}{\percent} to \SI{14.4}{\percent} and Interrupting scheduling \SI{13.3}{\percent} to \SI{18.9}{\percent} of \coo\ emissions.

Experiments that make use of the interruptibility of machine learning jobs are improving the achieved carbon savings by 24.2 to \SI{36.6}{\percent} for Germany, Great Britain, and France, and even by \SI{131.2}{\percent} for California. \autoref{fig:ml_jobs} shows the number of active jobs during an example period, demonstrating how Interrupting scheduling better exploits the daily fluctuation in carbon intensity than Non-Interrupting scheduling.

The additional flexibility enabled by semi-weekly scheduling causes the carbon savings to at least double across all regions, compared to experiments subject to the Next Workday constraint.
\autoref{fig:ml_constraints} depicts how semi-weekly constraint allows the scheduler to shift even more workload towards the weekend to avoid times of high carbon intensity. Moreover, also in the Monday to Thursday period, the emission rates are significantly lower than under the Next Workday constraint.

\begin{figure}[b]
    \includegraphics[width=0.9\linewidth]{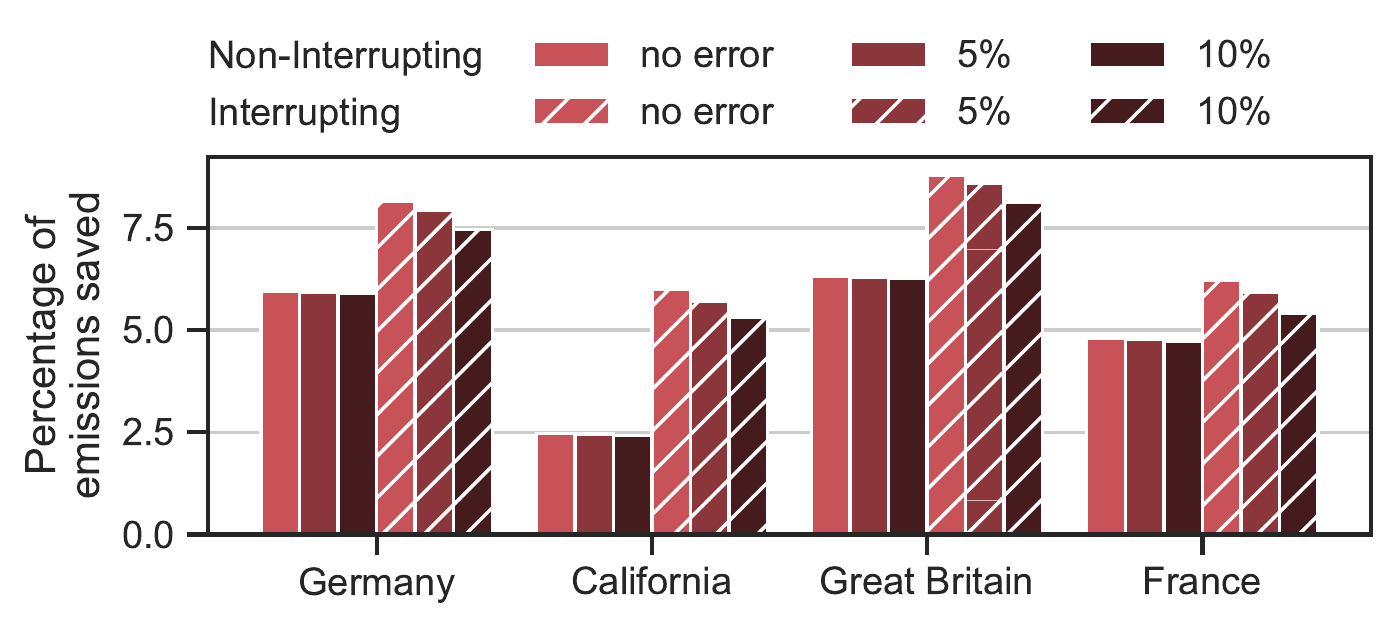}
	\caption{Influence of different forecast errors on carbon savings in the Next Workday constraint scenario.}
    	\label{fig:errors}
\end{figure}

\autoref{fig:errors} displays the effect of 5\,\% and 10\,\% forecast errors on the Next Workday constraint scenario.
While the savings for Non-Interrupting scheduling were almost the same independently from the applied errors, the Non-Interrupting scheduling highly benefits from low forecast errors.
Findings for the Semi-Weekly scenario were equivalent.

\subsubsection{Discussion}\label{sec:ml_discussion}

The experiments support our findings from \autoref{sec:analysis}: Shifting workloads towards nights and weekends, is a meaningful approach to consume cleaner energy. Even under time constraints that are not interfering with regular working hours, carbon savings of around \SI{5}{\percent} are possible. With more relaxed time constraints, results improve substantially. In practice this could be implemented by letting users define the date and time until which results are actually required.

Exploiting the interruptibility of workloads, proved to be an effective strategy to further reduce emissions. 
Future PaaS (Platform as a Service) and middleware systems should consider using snapshots not only for fault tolerance and possible evictions, but also for carbon-aware temporal load shifting.
Forecasts with error had almost no impact on the results of the Non-Interrupting strategy but considerable impact on Interrupting scheduling.
This is because Non-Interrupting scheduling optimizes for the lowest mean carbon intensity over entire intervals, and therefore is especially robust against noise in the forecasts.
Interrupting scheduling is more susceptible to optimize for negative spikes, however, even with 10\,\% forecast errors, it always outperforms Non-Interrupting scheduling.

To conclude, we observe that experiments exploiting the interruptibility of jobs at semi-weekly scheduling are the most successful.  
Since, according to the data from \cite{Karras_StyleGAN2ADA_2020}, a job consumes \SI{2036}{W} of power, in absolute numbers such a scheduling would have reduced the carbon emissions of the machine learning project by \SI{8.9}{t} if executed in Germany and \SI{6.3}{t} if executed in California or Great Britain. 
Although France has a very low mean carbon intensity already, savings of \SI{1.2}{t} were achieved. For comparison, the per capita emissions in Germany and Great Britain in 2019 were \SI{8.4}{t} and \SI{5.5}{t}, respectively \cite{OurWorldInData_C02_2020}.

\subsection{Limitations}

In our experiments, we did not consider any resource constraints, such as the available computational capacity at a given time.
While this is a reasonable assumption for Scenario I, in Scenario II there probably was a maximum number of GPUs available to the team.
However, the number of active jobs in the scheduling experiments did never exceed the maximum number of active jobs of the baseline experiment by more than \SI{42}{\percent} (64 compared to 45), which suggests that no unrealistic consolidation of workload took place.

Furthermore, forecast errors were simulated by applying uniform random noise on the actually observed carbon intensity. In reality, however, prediction errors are not uniform and also correlated. Errors grow with increasing forecast length, as well as during times with high variability such as daylight hours. Realistic forecasts may over- or underestimate the actual carbon intensity for multiple consecutive timestamps when relying, for example, on faulty weather forecasts.
Because of this, the validity of our forecast error analyses are limited. A more thorough analysis applying actual forecasts in different regions would be necessary to answer important questions such as how good a forecast should be to actually request a rescheduling.

\subsection{Implications}

This section summarizes implications and recommendations for the future design of services, schedulers, and middleware that emerge from our evaluation.

\subsubsection{Cloud and Service Providers}
To exploit fluctuations in carbon intensity, providers should generally encourage users to design their workloads to be temporally flexible and/or interruptible and to declare them as such. 
For example, preemtive VMs (also marketed as Spot VMs/Instances) are already available across many cloud providers, offering resources at a low cost with the goal to shape load in a way beneficial to the cloud operator.
As carbon pricing mechanisms may soon account for a considerable fraction of electricity costs \cite{WorldBank_CarbonPricing_2020}, this approach can also become profitable for carbon-aware load shaping.
However, as carbon intensity characteristics and carbon pricing mechanisms vary highly from region to region, the usefulness may be limited to certain locations and has to be re-evaluated on a regular basis.

Besides direct financial incentives, service providers can also incorporate knowledge about carbon intensity patterns and the associated costs into their SLA design.
For example, providing execution time windows (e.g. nightly) instead of exact times (e.g. every day at 1:00\,am) for certain services increases the temporal flexibility of workloads and, hence, the carbon saving potential.
Again, the data center's region plays a major role in the potential savings and has to be considered.

\subsubsection{Schedulers and Middleware}
Qualitative forecasts of carbon intensity and workload are a core component of any carbon-aware scheduler.
Luckily, short-term carbon intensity forecasts can often be predicted with high accuracy \cite{Leerbeck_TomorrowCo2Forecasting_2020, Bokde_ShortTermCO2ForecastingElectricityMarketScheduling_2021}; the same applies to many data center workloads, for example, in the domain of distributed stream processing \cite{Gontarska_TSFonDSP_2021}.
Besides, our research shows that the performance of carbon-aware schedulers highly depends on additional information about the workloads such as their temporal constraints, expected duration, and interruptibility.

Middleware systems can play an important role in providing this information to schedulers.
On the one hand, they should offer interfaces that allow different types of applications to conveniently declare temporal constraints and other properties of workloads programatically.
On the other hand, they can also feature automatic detection of certain characteristics.
For instance, systems that profile the time required to stop and resume a workload can automatically label it as interruptible or non-interruptible.
Likewise, temporal constraints could be derived by software that, for example, knows the dependency graph of tasks.

\section{Related Work}\label{sec:rw}
This section surveys related work in the field of renewable-aware workload scheduling, temporal workload shifting in the context of data center demand response, and grid carbon intensity forecasting.

\subsection{Renewable-Aware Scheduling}

Shaping data center load based on the availability of renewable energy has been a research topic for more than a decade \cite{Stewart_SomeJoulesAreMorePrecious_2009, Akoush_FreeLunch_2011, Zhang_GreenWare_2011, Goiri_GreenSlot_2011}, with a large fraction of the literature being from the early 2010s.
However, most methods focus on the integration and utilization of on-site and off-side renewable energy installations \cite{Fridgen_NotAllDoomAndGloom_2021, Beldiceanu_EnergyProportionalCloudsRenewableEnergy_2017, Akoush_FreeLunch_2011, Zhang_GreenWare_2011, Aksanli_GreenEnergyPredictionScheduleBatchServiceJobs_2011, Goiri_MatchingRenewableEnergyGreenDatacenters_2015, Liu_RenewableCoolingAwareWorkloadManagement_2012, Goiri_GreenHadoop_2012, Goiri_ParasolAndGreenSwitch_2013, Liu_RenewableCoolingAwareWorkloadManagement_2012, Li_iSwitch_2012, Dupont_ApplicationControllerForOptimizingRenewableEnergy_2015} and only few consider the carbon intensity of energy consumed from the power grid  \cite{Radovanovic_Google_2021, Zheng_MitigatingCarbonLoadMigration_2020, Zhou_CarbonAwareLoadBalancingGeoDistributedCloudServices_2013, Moghaddam_CarbonAwareDistributedCloudGenetic_2014}. 
Many approaches optimize for green energy by utilizing geo-distributed load migration, which is especially promising if data centers are being located in different hemispheres and time zones.
Free Lunch \cite{Akoush_FreeLunch_2011} and GreenWare \cite{Zhang_GreenWare_2011} are prominent examples of methods that reduce the amount of "wasted" renewable energy produced on-site by distributing workload among data centers, for example by virtual machine migration based on weather conditions.
Other approaches use geo-distributed workload shifting in order to directly consume energy with lower carbon intensity \cite{Zheng_MitigatingCarbonLoadMigration_2020, Zhou_CarbonAwareLoadBalancingGeoDistributedCloudServices_2013, Moghaddam_CarbonAwareDistributedCloudGenetic_2014}.

When considering renewable-aware scheduling within single data centers, the current literature focuses on the integration of renewable energy sources and does not consider the potential reduction of carbon intensity on the public grid.
For example, Aksanli et al. \cite{Aksanli_GreenEnergyPredictionScheduleBatchServiceJobs_2011} schedule workloads by utilizing short term prediction of solar and wind energy production. 
GreenSlot \cite{Goiri_MatchingRenewableEnergyGreenDatacenters_2015} schedules batch jobs that are executed in data centers with access to on-site solar energy generation by predicting the hourly availability of solar energy two days in advance.
Similarly, further approaches for renewable-aware schedulers \cite{Liu_RenewableCoolingAwareWorkloadManagement_2012, Goiri_GreenHadoop_2012, Goiri_ParasolAndGreenSwitch_2013} and works that consider the problem from a modeling \cite{Li_iSwitch_2012} or user \cite{Dupont_ApplicationControllerForOptimizingRenewableEnergy_2015} perspective, do not consider the carbon intensity of the power grids, neither do related surveys and reviews \cite{Oro_RenewableEnergyIntegrationReview_2015, KhosraviBuyya_SotaEnergyCarbonAwareDCs_2017, Shuja_SustainableCloudDataCenterSurvey_2016}.

Although Cappiello et al. already identified temporal shifting as a strategy to reduce emissions in cloud applications in 2015~\cite{Cappiello_CO2AwareAdaptationStrategies_2016}, the first and only work utilizing this technique to date is Google’s CICS~\cite{Radovanovic_Google_2021}.
They proactively shape compute load based on current and predicted power grid conditions and achieve power consumption drops of 1-2\,\% at times with the highest carbon intensity.
However, no information on the impact of CICS in different regions is provided.

\subsection{Demand Response in Data Centers}

Demand response and demand-side management describe the adjustment of power usage by end-consumers during times when the power grid is stressed to capacity.
The goal of demand response is to reduce peak electricity demand and, hence, to increase the stability of the power grid.
In practice, this is usually achieved by providing financial incentives to consumers \cite{EuropeanCommission_DemandResponse_2013}. From an operators perspective demand response programs are therefore mainly an opportunity to reduce costs, not emissions.

Data centers have been identified as a promising industry for demand response because they consume large amounts of energy while being flexible due to their automated nature~\cite{Shoreh_SurveyIndustrialApplicationsDemandRespons_2016}.
An in depth field study of data center demand response by Lawrence Berkeley National Laboratories (LBNL) \cite{Ghatikar_DemandResponseOpportunitiesFieldStudies_2012} concludes that postponing computational load is an important demand response strategy next to load migration, shutting down or idling IT equipment, adjusting cooling, and adjusting building properties like lighting. 
Several works have since investigated this flexibility \cite{Basmadjian_FlexibilityBasedEnergyDemandManagementCaseStudyCloud_2019, Vasques_DemandResponseSmallMediumDCs_2019, Klingert_MappingBusinessTypes_2018, Klingert_SpinningGoldFromStraw_2020, Wierman_OpportunitiesChallengesDemandResponse_2014} and have proposed solutions to exploit it \cite{Cioara_OptimizedFlexibilitySmartDemandResponse_2018, Klingert_SpinningGoldFromStraw_2020, Cupelli_ControlStrategyForParticipationInDR_2018, Cioara_ExploitingFlexibilitySmartCitiesBusinessScenarios_2019, Chen_DataCenterGridStabilizer_2014, Yao_TwoTimeScaleApproachDelayTolerantWorkloads_2012}. 
Existing literature also considers demand response in conjunction with local power generation \cite{Liu_DemandResponseCoincidentPeak_2013} or focuses on directly forecasting energy flexibility \cite{Vesa_EnergyFlexibilityPredictionDemandResponse_2020}.

Data center demand response is working towards adapting the power demand profile of data centers.
However, current efforts focus on power grid stability and usually optimize for cost effectiveness in incentive-based or price-based scenarios.
Contrarily, our aim is to evaluate the potential of temporal workload shifting in regards to carbon savings.

\subsection{Carbon Intensity Forecasts}\label{sec:rw_forecasts}

In recent years, it has become increasingly popular to utilize carbon intensity forecasts to adaptively control power usage, for example in residential heating \cite{Pean_CarbonBasedEnergyFlexibilityResidentialHeating_2019, VoglerFinck_CarbonFootprintHouseHeating_2018, Pean_EnvironmentalImpactDREnergyFlexibleBuildings_2018} or smart charging battery electric vehicles \cite{Huber_CarbonEfficientSmartChargingMarginalEmissionForecasts_2020}. 
However, while there are plenty of long-term forecasting models on \coo{} emissions of countries or industrial sectors, comparably little research exists on predicting short-term grid carbon intensity.

The most prominent supplier of carbon intensity data is Tomorrow's \emph{electricityMap} that also provides the data for CICS.
While their methodologies on real-time consumption-based carbon accounting \cite{Tranberg_CarbonAccounting_2019} as well as short-term carbon intensity forecasting for average and marginal emissions \cite{Leerbeck_TomorrowCo2Forecasting_2020} are publicly available, their data is only to a certain degree.
An open carbon intensity forecast is provided by the National Grid ESO~\cite{NationalGridESO_ForecastMethodology_2021}, a power grid operator in Great Britain. 
Their so called \emph{Carbon Intensity API} provides 96 hour forecasts for different regions in Great Britain based on a rolling-window linear regression model and uses a methodology for computing carbon intensity that is similar to ours. However, their forecasting model is not open source and relies on non-publicly available weather data, meaning it cannot be transferred to other regions.
Lowry \cite{Lowry_DayAheadForecastingCarbonIntensityHVAC_2018} uses autoregressive integrated moving average (ARIMA) and neural network models for day-ahead forecasting of grid carbon intensity in order to control heating, ventilation, and air conditioning systems.
Lastly, Bodke et al. \cite{Bokde_ShortTermCO2ForecastingElectricityMarketScheduling_2021} use a decomposition approach and forecast the grid carbon intensity of regions within Europe via statistical methods.

\section{Conclusion}\label{sec:conclusion}

This paper examines the potential of temporally shifting computational workloads in data centers with the goal to consume cleaner energy from the public power grid.
We provide an overview on characteristics of shiftable workloads and analyze the regional carbon intensity of Germany, California, Great Britain, and France over the year 2020. 
Our findings suggest that short-term shifting potential is often high before sunrise in countries with a lot of solar power, and in the evening hours, because most countries throttle their fossil fuel power stations at night. 
Moreover, shifting delay-tolerant workloads towards weekends can result in more than \SI{20}{\percent} savings in most regions.
The experimental evaluation supports our analytical findings and demonstrates that the highest savings can be achieved when relaxing time constraints and actively exploiting the interruptibility of workloads during scheduling.
For example, shifting workloads whose results are not needed by the next working day can already reduce emissions by over 5\,\% across all regions.

Future work will address the development and evaluation of schedulers that take advantage of the findings in this paper.
To this end, we hope that our simulator and published datasets will prove to be useful tools for exploring new approaches in this domain.
In particular, we want to use them to research on the combination of temporal and geo-distributed scheduling, which has received little attention to date.

\begin{acks}
We would like to thank all Middleware reviewers for their valuable comments and suggestions.
We also express our sincere thanks to Martin Schellenberger for his insights that helped to shape this work.
This research was supported by the German Academic Exchange Service (DAAD) as ide3a and the German Ministry for~Education and Research (BMBF) as \mbox{BIFOLD} (research grant 01IS18025A).
\end{acks}

\balance

\bibliographystyle{ACM-Reference-Format}
\bibliography{bibliography}


\end{document}